%% file: bracl.tex
\documentclass[11pt, a4paper]{article}

\usepackage{tikz}
\usepackage{amsthm}
\usepackage{amsmath}
\usepackage{amssymb}
\usepackage{amsfonts}
\usepackage{latexsym}
\usepackage{booktabs}
\usepackage{multirow}
\usepackage{graphicx}
\usepackage{natbib}
\usepackage{pdfpages}
\usepackage{algorithm}
\usepackage{algpseudocode}
\usepackage[left = 2.5cm, right = 2.5cm, top = 2.7cm, bottom = 3cm]{geometry}
\usepackage[colorlinks,citecolor=black,urlcolor=black]{hyperref}

\def\trace{{\mathop{\rm trace}}}
\def\expect{{\mathop{\rm E}}}

\def\det{{\mathop{\rm det}}}

\newtheoremstyle{example}
{3pt} 
{3pt} 
{} 
{0\parindent} 
{\bf}
{:} 
{.5em} 
{} 
\newtheoremstyle{theorem}
{3pt} 
{3pt} 
{\em} 
{0\parindent} 
{\bf}
{:} 
{.5em} 
{} 
\theoremstyle{example} \newtheorem{example}{Example}
\theoremstyle{theorem} 
\theoremstyle{lemma} 
\theoremstyle{corollary}

\author{Ioannis Kosmidis \\ \texttt{ioannis.kosmidis@warwick.ac.uk} \bigskip \\
  Department of Statistics, University of Warwick\\
  Coventry CV4 7AL, UK
}

\title{Mean and median bias reduction: A concise review and
  application to adjacent-categories logit models}

\begin{document}

\maketitle

\begin{abstract}
  The estimation of categorical response models using bias-reducing
  adjusted score equations has seen extensive theoretical research and
  applied use. The resulting estimates have been found to have
  superior frequentist properties to what maximum likelihood generally
  delivers and to be finite, even in cases where the maximum
  likelihood estimates are infinite. We briefly review mean and median
  bias reduction of maximum likelihood estimates via adjusted score
  equations in an illustration-driven way, and discuss their
  particular equivariance properties under parameter transformations.
  We then apply mean and median bias reduction to adjacent-categories
  logit models for ordinal responses. We show how ready bias reduction
  procedures for Poisson log-linear models can be used for mean and
  median bias reduction in adjacent-categories logit models with
  proportional odds and mean bias-reduced estimation in models with
  non-proportional odds. As in binomial logistic regression, the
  reduced-bias estimates are found to be finite even in cases where
  the maximum likelihood estimates are infinite. We also use the
  approximation of the bias of transformations of mean bias-reduced
  estimators to correct for the mean bias of model-based ordinal
  superiority measures. All developments are motivated and
  illustrated using real-data case studies and simulations. 
\end{abstract}

\section{Overview}

The first part of this chapter provides an example-driven, concise
review of the developments in a fast growing body of literature about
mean and median bias reduction (BR) in parametric estimation via
adjusted score equations; see \citet{firth:1993} for mean BR (mBR)
and \citet{kenne+salvan+sartori:2017} for median BR (mdBR).
Particular focus is placed on how these methods can be used as a
remedy for the numerical and inferential consequences of boundary
maximum likelihood (ML) estimates in categorical response models,
which are illustrated in
Section~\ref{sec:boundary}. Sections~\ref{sec:meanBR} and
\ref{sec:medianBR} describe how the mean and median bias of the ML
estimator can be reduced in general parametric models through the
appropriate adjustment of the gradient of the
log-likelihood. Section~\ref{sec:inference} discusses the validity of
inference when the ML estimates are replaced by mBR or mdBR estimates
in standard first-order procedures. Section~\ref{sec:bias_trans} takes
a close look at the equivariance properties of mBR and mdBR estimators
under transformation of the model parameters. We also present an
approximation of the bias of general transformations of mBR
estimators, which can be used to correct for the bias of
transformations of the model parameters using only the mBR
estimates, the second derivatives of the transformation, and the
expected information matrix. The bias approximation is used to get mBR
estimates of odds ratios from mBR estimates of regression coefficients
in logistic regression models.

The second part of this chapter uses the results from the first to
develop, for the first time, mBR and mdBR procedures for
adjacent-categories logit (ACL) models for ordinal responses
\citep[see, for example,][Chapter~4 for an
introduction]{agresti:2010}. Section~\ref{sec:acl_models} reviews the
proportional odds (PO) and non-proportional odds (NPO) versions of the
ACL models, and their key properties, including their equivalence to
baseline-category logit (BCL) models, and discusses how that
equivalence can be exploited for ML estimation. A real-data case study
is used to illustrate that boundary estimates can also cause numerical and
inferential issues for ACL models. Section~\ref{sec:acl_meanBR}
then details how and when the equivariance properties of mBR and mdBR,
and implementations of the latter for BCL models, can be used for mBR
and mdBR for the PO and NPO versions of ACL models. Finally,
Section~\ref{sec:meanBR_superiority} details how the mBR estimates can
be used for the explicit correction of the estimates of ordinal
superiority summaries.

\section{Boundary estimates in categorical response models}
\label{sec:boundary}
It is well known that ML estimation of regression models with
categorical responses may result in estimates on the boundary of the
parameter space.  The data patterns that result in boundary estimates
in general multinomial logistic regression models (also known as
baseline category models models; see
\citealp[Section~7.1]{agresti:2002}) have been studied extensively and
are completely characterized. For a range of binomial regression
models, \citet{silvapulle:1981} proves that a certain degree of
``overlap'' on the data is a necessary and sufficient condition for
the ML estimates to have finite values. \citet{albert+anderson:1984}
enrich the arguments in \citet{silvapulle:1981} generalizing the
results in the case of baseline-category logit (BCL) models for
nominal responses. In particular, \citet{albert+anderson:1984}
categorize the possible configurations for the sample points into
complete separation, quasi-complete separation, and overlap, and then
show that separation is necessary and sufficient for the ML estimate
to have at least one infinite-valued component. Geometric
representations of (quasi-)complete separation for binomial logistic
regression --- when the ACL and BCL models reduce to exactly the same
form --- are given in \citet[Figure 1]{albert+anderson:1984}, and for
multinomial responses in \citet[Figure 1]{lesaffre+albert:1989}.

\begin{example}
  \label{ex:separation}
  {\bf Separation in logistic regression} A simple illustration of a
  completely separated data set is shown in
  Figure~\ref{fig:logist_separation}. The data consists of $100$
  realizations of two continuous covariates $x_2$ and $x_3$, and a
  response $y$ that ends up being $0$ whenever $x_2 + 2 x_3 > 0$. ML
  estimation of the logistic regression model with
  $\log \{ \pi / (1 - \pi) \} = \beta_1 + \beta_2 x_2 + \beta_3 x_3$,
  where $\pi$ is the probability of observing $y = 1$ given $x$,
  results in the estimated logistic discriminant line in
  Figure~\ref{fig:logist_separation}, with the log-likelihood
  attaining its global maximum value of $0$, and the fitted value $0$
  being assigned to all observations with $y = 0$, and $1$ to the
  rest. The \texttt{detectseparation} R package
  \citep{detectseparation} that implements the methods in the
  unpublished PhD thesis by \citet{konis:2007} can be used to show
  that the ML estimates of $\beta_1$, $\beta_2$ and $\beta_3$ are
  $-\infty$, $+\infty$ and $+\infty$ respectively.

\begin{table}[t]
  \caption{ML, mBR and mdBR estimates for the logistic regression
    model in Example~\ref{ex:separation}. The estimated standard
    errors (S.E.) are based on the expected information matrix at the
    estimates. The $z$-statistic is computed as estimate over
    estimated S.E., and the $p$-value is computed as
    $2\min(\Phi($z$), 1- \Phi($z$))$, where $\Phi(\cdot)$ is the cumulative distribution function
    of the standard Normal distribution.}
\begin{center}
  \begin{tabular}{lrrrr}
    \toprule
    \multicolumn{1}{c}{Parameter} & \multicolumn{1}{c}{Estimate} & \multicolumn{1}{c}{Estimated S.E.} & \multicolumn{1}{c}{$z$-statistic} & \multicolumn{1}{c}{$p$-value} \\ \midrule
    \multicolumn{5}{c}{Maximum likelihood} \\ \midrule
    $\beta_1$ & -22.397 & 13879.616 & -0.002 & 0.999 \\
    $\beta_2$ & 62.578 & 20968.761 &  0.003 & 0.998 \\
    $\beta_3$ & 132.228 & 44964.541 &  0.003 & 0.998 \\ \midrule
    \multicolumn{5}{c}{Mean bias reduction} \\ \midrule
    $\beta_1$ & -2.001 & 1.552 & -1.289 & 0.197 \\ 
    $\beta_2$ & 5.266 & 1.997 &  2.637 & 0.008 \\
    $\beta_3$ & 11.166 & 3.984 &  2.803 & 0.005 \\ \midrule
    \multicolumn{5}{c}{Median bias reduction} \\ \midrule
    $\beta_1$ & -2.583 & 1.984 & -1.302 & 0.193 \\ 
    $\beta_2$ & 6.325 & 2.628 &  2.406 & 0.016 \\
    $\beta_3$ & 13.321 & 5.293 &  2.517 & 0.012 \\ \bottomrule
  \end{tabular}
\end{center}
\label{tab:logist_separation}
\end{table}

\begin{figure}[t!]
  \begin{center}
    \includegraphics[width = 0.7\textwidth]{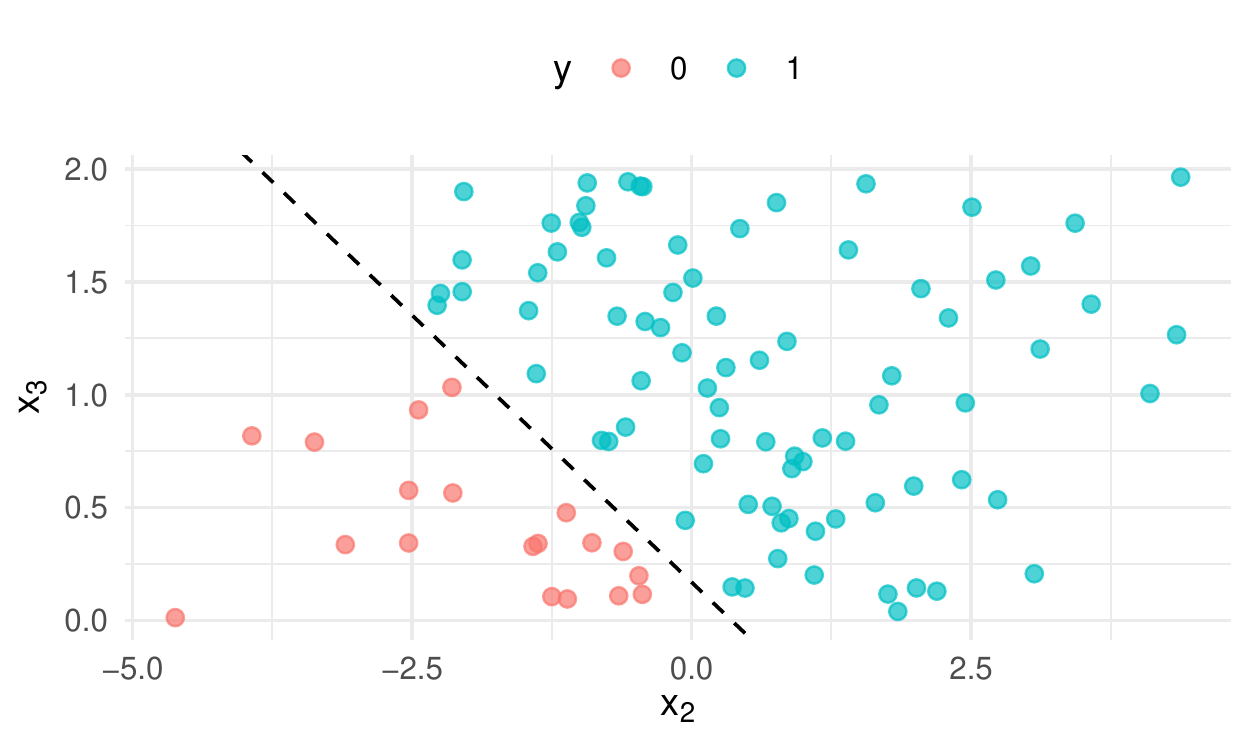}
  \end{center}
  \caption{The data described in Example~\ref{ex:separation}. The
    dashed line is the line $0 = \hat\beta_2 x_2 + \hat\beta_3 x_3$,
    where the fitted probabilities are all $0.5$.}
  \label{fig:logist_separation}
\end{figure}

While there is no ambiguity in reporting infinite estimates, estimates
on the boundary of the parameter space can i) cause numerical
instabilities to fitting procedures, ii) lead to misleading output
when estimation is based on iterative procedures with a stopping
criterion, and more importantly, iii) cause havoc to asymptotic
inferential procedures, and especially to the ones that depend on
estimates of the standard error of the estimators (for example, Wald
tests and related confidence intervals), oftentimes leading to wrong
inferences. For example, the ML estimates in
Table~\ref{tab:logist_separation} have been obtained using the
\texttt{glm()} function in R \citep{rproject}. Despite the fact that
the ML estimates for $\beta_1$, $\beta_2$ and $\beta_3$ are in reality
infinite, the stopping criteria of the fitting procedure that
\texttt{glm()} implements are met for finite values of the parameters,
which are returned. The reported estimated standard errors are also
finite and substantially larger than the estimates. This results in
small, in absolute value, $z$-statistics, and hence no evidence against
the individual hypotheses $\beta_2 = 0$ and $\beta_3 = 0$; one would
expect at least some evidence against the hypotheses given that the
value of the response has been fully determined from the values of
$x_2$ and $x_3$.
\end{example}

One way to circumvent the numerical and inferential issues associated
with boundary ML estimates is to replace ML with an alternative
estimation method that i) has comparable or sometimes better
asymptotic properties than the ML estimator generally does, and ii)
tends to result or results in estimates away from the boundary of the
parameter space. Popular examples of such alternative estimation
methods are the mean bias-reducing adjusted score functions approach
in \citet{firth:1993}, and the median bias-reducing adjusted score
functions approach in \citet{kenne+salvan+sartori:2017}, which we
briefly review in Section~\ref{sec:meanBR},
Section~\ref{sec:medianBR}, and Section~\ref{sec:bias_trans}.

\section{Mean bias reduction} 
\label{sec:meanBR}

Let $\ell(\theta)$ be the log-likelihood about a parameter vector
$\theta$ with $\theta \in \Re^v$. Assuming that the model at hand is
appropriate, then under fairly general regularity conditions about
the model, the ML estimator $\hat\theta = \arg \max \ell(\theta)$ has
mean bias $\expect_\theta(\hat\theta - \theta) = O(N^{-1})$, where $N$
is a measure of information about $\theta$, usually --- but not
necessarily --- the sample size.

If $S(\theta) = \nabla \ell(\theta)$, \citet{firth:1993} shows that we
can define an alternative estimator $\theta^*$ with mean bias
$\expect_\theta(\theta^* - \theta) = O(N^{-2})$, which is
asymptotically smaller than the bias of $\hat\theta$, as the solution
of
\begin{equation}
  \label{eq:meanbr}
  S(\theta) + A(\theta) = 0_v \, ,
\end{equation}
where
\[
  A_t(\theta) = \frac{1}{2} \trace\left[i(\theta)^{-1} \left\{
      P_t(\theta) + Q_t(\theta)\right\}\right] \quad (t= 1, \ldots, v)\, .
\]
In the above expression,
$P_t(\theta) = \expect_\theta(S(\theta) S(\theta)^\top
  S_t(\theta))$ and
$Q_t(\theta) = -\expect_\theta(j(\theta) S_t(\theta))$,
and $j(\theta) = -\nabla \nabla^\top \ell(\theta)$ and
$i(\theta) = \expect_\theta(S(\theta) S(\theta)^\top)$ are the
observed and expected information matrix about $\theta$, respectively,
with all expectations taken with respect to the model.

Mean bias reduction has been found to result in estimates away from
the boundary of the parameter space in a range of categorical data
models; see, for example, \citet{firth:1993} and
\citet{heinze+schemper:2002} for binomial logistic regression;
\citet{mehrabi+matthews:1995} for the estimation of simple
complementary log–log-models; \citet[Section~6]{kosmidis+firth:2009}
for row-column association models; \citet{bull:02},
\citet{kosmidis+firth:2011}, and
\citet[Section~6]{kosmidis+kennepagui+sartori:2020} for BCL models;
and \citet{kosmidis:2014} for cumulative link models.

If $\theta$ is the canonical parameter of a full exponential family
\citep[see][Chapter 5]{pace+salvan:1997}, like in binomial and
multinomial logistic regression, then $j(\theta) = i(\theta)$ and
$j(\theta)$ does not depend on the stochastic part of the
model. Hence, $Q_t(\theta) = 0_{v \times v}$, where $0_{v \times v}$
is a $v \times v$ matrix of zeros, and some algebra
\citep[see][Section~3]{firth:1993} gives that the solution of the
mean bias-reducing adjusted score equations~\eqref{eq:meanbr} is
equivalent to the maximization of the penalized log-likelihood
\begin{equation}
  \label{eq:penloglik}
  \ell(\theta) + \frac{1}{2} \log\det \{ i(\theta) \} \, ,
\end{equation}
where the penalty is the logarithm of the Jeffreys prior. Recent work
by \citet{kosmidis+firth:2021} considers the impact of penalized
likelihoods like~(\ref{eq:penloglik}) in the estimation of many
well-used binomial-response generalized linear models, including
logistic, probit, complementary log-log, and cauchit
regression. Among other results, \citet{kosmidis+firth:2021} prove
that maximizing the likelihood after penalizing it by arbitrary
positive powers of the Jeffreys prior always results in finite
estimates, and derive the shrinkage directions implied by the penalty.

\section{Median bias reduction}
\label{sec:medianBR}

The median bias-reducing adjusted score functions of
\citet{kenne+salvan+sartori:2017} is another method that has been
found to result in finite estimates in extensive simulation studies
with logistic regression and BCL models
\citep[see][Section~6]{kosmidis+kennepagui+sartori:2020}) and with
cumulative link models \citep{gioia+kennepagui+salvan:2021}.

The ML estimator generally has median bias
$P(\hat\theta_t \le \theta_t) = 1/2 +
O(N^{-1/2})$. \citet{kenne+salvan+sartori:2017} show that we can
define an alternative estimator $\theta^\dagger$ with
$P(\theta_t^\dagger \le \theta_t) = 1/2 + O(N^{-3/2})$, which is
asymptotically closer to $1/2$ than the median bias of $\hat\theta$,
as the solution of
\begin{equation}
  \label{eq:medianbr}
  S(\theta) + A(\theta) - i(\theta) F(\theta) = 0_v \, .
\end{equation}
In the above expression,
$F_t(\theta) = [i(\theta)^{-1}]_t^\top \tilde{F}_t(\theta)$, with
\[
  \tilde{F}_{tu}(\theta) =  \trace\left[\tilde{i}_u(\theta) \left\{
      \frac{1}{3}P_t(\theta) + \frac{1}{2}Q_t(\theta)\right\}\right] \quad (t= 1, \ldots, g) \, ,
\]
and
$\tilde{i}_u(\theta) = [i(\theta)^{-1}]_u[i(\theta)^{-1}]_u^\top /
[i(\theta)^{-1}]_{uu}$ $(u = 1, \ldots, v)$, where $A_u$ and $A_{tu}$
denote the $u$th column and $(t, u)$th element of a matrix $A$.

When $j(\theta) = i(\theta)$, expression~(\ref{eq:medianbr})
simplifies in a similar manner as expression~(\ref{eq:meanbr})
does. In fact, for one-parameter models ($v = 1$) that are exponential
families in canonical parameterization, it can be shown that mdBR is
formally equivalent to the maximization of
$\ell(\theta) + \log\det\{i(\theta)\} / 6$
\citep[see][Section~2.1]{kenne+salvan+sartori:2017}. However, mdBR
has no penalized likelihood interpretation for $v > 1$.

\section{Inference with mean and median bias reduction}
\label{sec:inference}

According to the results in \citet{firth:1993} and
\citet{kenne+salvan+sartori:2017}, both $\theta^*$ and
$\theta^\dagger$ have the same asymptotic distribution as the ML
estimator generally does, and hence are asymptotically efficient. Therefore,
the distribution of those estimators for finite samples can be
approximated by a Normal with mean $\theta$ and variance-covariance
matrix $\{i(\theta)\}^{-1}$. The derivation of this result relies on
the fact that both the adjustments $A(\theta)$ and
$A(\theta) - i(\theta) F(\theta)$ to the score functions for mBR and
mdBR in~(\ref{eq:meanbr}) and~(\ref{eq:medianbr}), respectively, are
of order $O(1)$ as $N \to \infty$. Hence, the score function
$S(\theta)$, which is $O_p(\sqrt{N})$, dominates the adjustments as
information increases. The implication is that standard errors for the
components of $\theta^*$ and $\theta^\dagger$ can be computed exactly
as for the ML estimator, using the square roots of the diagonal
elements of $\{i(\theta)\}^{-1}$ of $\{j(\theta)\}^{-1}$ at the
estimates. Furthermore, first-order inferences, like standard Wald
tests and Wald-type confidence intervals and regions are constructed
in a plugin fashion, by replacing the ML estimates with the mBR or
mdBR estimates in the usual procedures in standard software.

\begin{example}
  \label{ex:separation2}
  {\bf Separation in logistic regression (continued)} Continuing from
  Example~\ref{ex:separation}, Table~\ref{tab:logist_separation}
  provides the estimates of $\beta_1$, $\beta_2$ and $\beta_3$ from
  mBR and mdBR. The estimates have been computed using the default
  arguments of the \texttt{brglm\_fit()} method of the \texttt{brglm2}
  R package \citep{brglm2}. \texttt{brglm\_fit()} implements a variant
  of the quasi-Fisher scoring procedure
  \begin{equation}
    \label{eq:quasi_fisher}
    \theta^{(k + 1)} = \theta^{(k)} + \{i(\theta^{(k)})\}^{-1} U(\theta^{(k)}) \, ,
  \end{equation}
  where $U(\theta) := S(\theta) + A(\theta)$ if the intention is to
  compute the mean BR estimates, and
  $U(\theta) := S(\theta) + A(\theta) - i(\theta) F(\theta)$ if the
  intention is to compute the mdBR estimates; see
  \citet{kosmidis+kennepagui+sartori:2020} for details on the
  quasi-Fisher iterations and the form of the adjusted scores for mBR
  and mdBR in generalized linear models. Convergence has been rapid
  and \texttt{brglm\_fit()} reported no issues for either mBR or
  mdBR. Furthermore, the estimates and estimated standard errors
  appear to be finite. Note that the estimates and estimated standard
  errors from mBR are typically closer in absolute value to zero than
  those from mdBR.  Importantly, the $z$-statistics for $\beta_2$ and
  $\beta_3$ are all away from zero, and, in contrast to ML, both mBR
  and mdBR suggest at least some evidence against the individual
  hypothesis $\beta_2 = 0$ and $\beta_3 = 0$, which agrees with the
  fact that the value of the response has been fully determined from
  the values of $x_2$ and $x_3$.
\end{example}

\section{Bias reduction and parameter transformation}
\label{sec:bias_trans}

\subsection{Maximum likelihood estimation and general parameter transformations}

The ML estimator is equivariant in the sense that the ML estimator of
$g(\theta)$ is exactly $g(\hat\theta)$ for any one-to-one
transformation $g(\cdot)$. Hence, there is no need to maximize the
log-likelihood about $g(\theta)$ if the ML estimator of $\theta$ has
already been computed. In contrast, the mBR and mdBR estimators are
equivariant only for specific transformations $g(\cdot)$.

\subsection{Mean bias reduction and linear parameter transformations}
\label{sec:meanBR_trans}

The mBR estimator is equivariant under linear transformations for the
parameters, in the sense that the mBR estimator of $C\theta$ for a
known matrix $C$ is exactly $C\theta^*$. The same is not true for the
mdBR estimator.
 
For example, using Table~\ref{tab:logist_separation},
the mBR estimate of $\beta_2 - \beta_3$ in
Example~\ref{ex:separation2} is simply $5.266 - 11.166 = -5.9$. The
mdBR estimate, however, is not $6.325 - 13.321 = 6.996$, but rather
$-7.227$, which is obtained by reparameterizing the model in terms of
$\beta_2 - \beta_3$ and computing the mdBR estimate by
solving~(\ref{eq:medianbr}) in the new parameterization.

\subsection{Median bias reduction and component-wise parameter transformations}
\label{sec:medianBR_trans}

On the other hand, the mdBR estimator of
$(g_1(\theta_1), \ldots, g_v(\theta_v))^\top$ is
$(g_1(\theta_1^\dagger), \ldots, g_v(\theta_v^\dagger))^\top$ for any
set of one-to-one functions $g_1(\cdot), \ldots, g_v(\cdot)$. In other
words, the mdBR estimator is equivariant under component-wise
transformations. The same is not true for the mBR estimator. For
example, the mdBR estimate of the odds-ratio $\exp(\beta_2)$ in
Example~\ref{ex:separation2} is exactly $\exp(6.325)$, but
$\exp(5.266)$ is not an mBR estimate of $\exp(\beta_2)$.

\subsection{Mean bias reduction and general parameter transformations}

\citet{dicaterina+kosmidis:2019} show that there is a simple way to
derive the mean bias of $h(\theta^*)$ for any three-times
differentiable function $h: C \to D$, with $C \subset \Re^p$ and
$D \subset \Re$, where $\theta^*$ is an mBR estimator of $\theta$ with
$O(N^{-2})$ bias. In particular, \citet{dicaterina+kosmidis:2019} show
that the estimator $h(\theta^*)$ of $\zeta = h(\theta)$ has mean bias
\begin{equation}
  \label{eq:bias_trans}
  E(h(\theta^*) - h(\theta)) = \frac{1}{2} \trace \left\{ i(\theta)^{-1} \nabla\nabla^\top h(\theta) \right\} + O(N^{-2}) \, ,
\end{equation}
where $\nabla\nabla^\top h(\theta)$ is the hessian of $h(\cdot)$ at
$\theta$.  Note that for linear transformations,
$\nabla\nabla^\top h(\theta) = 0_{v \times v}$, and hence
$E(h(\theta^*) - h(\theta)) = O(N^{-2})$, which confirms the
discussion in Section~\ref{sec:meanBR_trans} that the mBR estimator is
exactly equivariant for linear transformations of the parameters. The
first term in the right-hand side of~(\ref{eq:bias_trans}) can be
evaluated at $\theta^*$ and be used to derive mean BR estimators of
$h(\theta)$, based only on $\hat\theta^*$, $i(\hat\theta^*)$, and
$\nabla\nabla^\top h(\theta^*)$. An obvious mean BR estimator
resulting from~(\ref{eq:bias_trans}) is
$h(\theta^*) - \trace \left\{ i(\theta^*)^{-1} \nabla\nabla^\top
  h(\theta^*) \right\} / 2$.

For example, consider the special case of estimation of the odds-ratio
$\exp(\beta_j)$ in Example~\ref{ex:separation}, which was estimated
using the equivariance properties of mdBR in
Section~\ref{sec:medianBR_trans}. Expression~(\ref{eq:bias_trans})
gives that the odds-ratio at the mBR estimator has
\begin{equation}
  \label{eq:bias_exp}
  E(\exp(\beta_j^*)) = \exp(\beta_j)\left[1 + \frac{1}{2} v_{jj}(\theta)\right] + O(N^{-2}) \, ,
\end{equation}
where $v_{jj}(\theta) = [i(\theta)^{-1}]_{jj}$. Hence, two mean BR estimators of $\zeta_j = \exp(\beta_j)$ with $O(N^{-2})$ bias are
\[
  \zeta_j^* = \exp(\beta_j^*) \left[ 1 - \frac{1}{2} v_{jj}(\theta^*) \right] \quad \text{and} \quad \zeta_j^{**} = \frac{\exp(\beta_j^*)}{1 + v_{jj}(\theta^*)/2} \, ,
\]
arising from subtracting an estimate of the bias at
$\theta := \theta^*$ from $\exp(\beta_j^*)$, and dividing
$\exp(\beta_j^*)$ by the correction factor $1 + v_{jj}(\theta^*)/2$
from the right-hand side of~(\ref{eq:bias_exp}), respectively. The
estimator $\zeta_j^{**}$ for the odds-ratio $\zeta_j$ has the
advantage of being always positive, while $\zeta_j^{*}$ takes negative
values if $v_{jj}(\theta^*) > 2$. For example, to the accuracy
reported in Table~\ref{tab:logist_separation},
$\zeta_2^* = \exp(5.266) (1 - 1.997^2 / 2) = -192.48$, which is
clearly nonsensical as an odds-ratio estimate. In contrast,
$\zeta_2^* = \exp(5.266) / (1 + 1.997^2 / 2) = 64.66$.  The
approximation
$\exp\{v_{jj}(\theta) / 2\} \approx 1 + v_{jj}(\theta)/2$ for small
$v_{jj}(\theta)$ can be used to show that the mean BR estimator
$\zeta_j^{**}$ closely relates to the mean BR estimator
$\zeta_j^{***} = \exp\{\beta_j^* - v_{jj}(\theta^*) / 2\}$ derived in
\citet{lyles+guo+greenland:2012}.

The discussion in Section~\ref{sec:inference} implies that estimated
standard errors for mBR estimators of transformed parameters
constructed on the basis of~(\ref{eq:bias_exp}) can be computed using
the delta method, as for the ML estimator.

\section{Adjacent-categories logit models}
\label{sec:acl_models}

\subsection{Proportional and non-proportional odds models}
\label{sec:prop-non-prop}

We now turn our attention in applying mBR and mdBR from
Section~\ref{sec:meanBR} and Section~\ref{sec:medianBR} to ACL models.

Adjacent-categories logit models \citep[see, for example,][Chapter~4
for an introduction]{agresti:2010} are a prominent family of regression
models for ordinal responses, where the local odds ratios of
consecutive categories of an ordinal response variable are linked with
linear combinations of parameters and explanatory variables. Suppose
that we observe realizations of $n$ independent random vectors of
frequencies $Y_1, \ldots, Y_n$, where
$Y_i = (Y_{i1}, \ldots, Y_{ik})^\top$ has a $k$-category multinomial
distribution with ordered categories $1 < 2 < \ldots < k$, total
$m_i = \sum_{j = 1}^k Y_{ij}$ and probability vector
$(\pi_{1}(x_i), \ldots, \pi_{k}(x_i))^\top$ with
$\sum_{j = 1}^k \pi_{j}(x_i) = 1$, where
$x_i = (x_{i1}, \ldots, x_{ip})^\top$ is a $p$-vector of covariate
values. An ACL model has
\begin{equation}
  \label{eq:po}
  \log \frac{\pi_{j}(x)}{\pi_{j+1}(x)} = \eta_{j}(x) \quad (j = 1, \ldots, k-1) \, ,
\end{equation}
where $\eta_{j}(x)$ is typically a linear combination of unknown model
parameters and a covariate vector $x$.

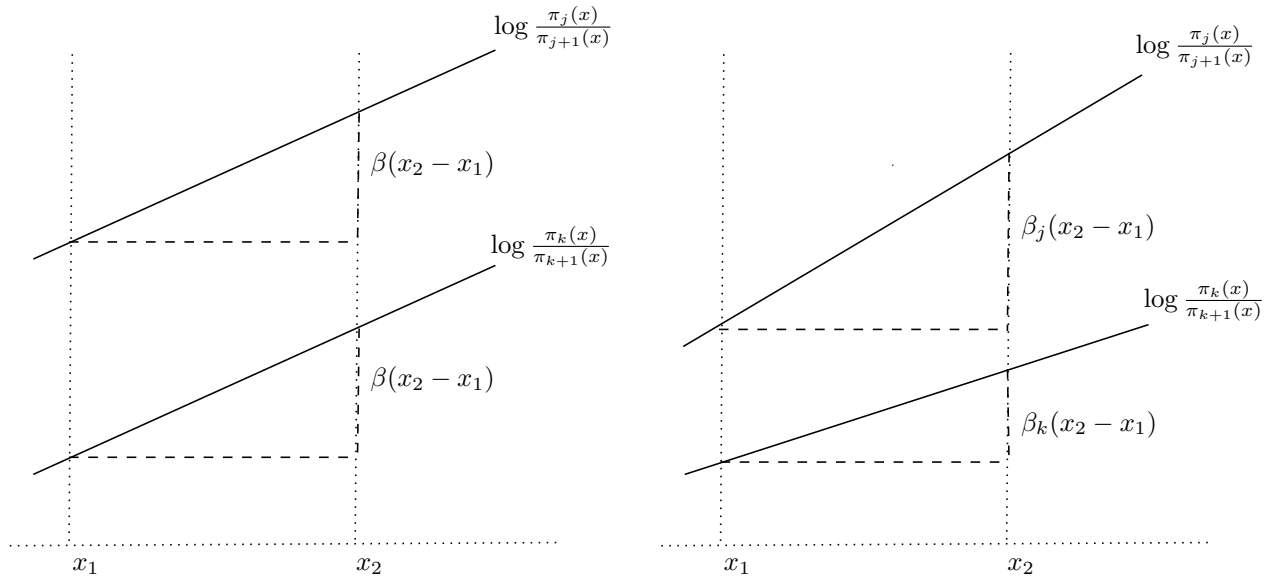
\begin{figure}[t]
  \caption{The adjacent-categories log-odds for categories $j$ and
    $k$, $j \ne k$, under the proportional odds (left) and the
    non-proportional odds (right) versions of the model, for
    $x \in \Re$. The probability for category $j$ at covariate value
    $x$ is denoted $\pi_j(x)$.}
  \begin{small}
    \input{figures/PO.tex}
  \end{small}
  \label{fig:po}
\end{figure}

The specification of $\eta_j(x)$ results in ACL models with particular
properties. The PO version of the ACL model has
\begin{equation}
  \label{eq:polp}
  \eta_{j}(x) = \alpha_j + \beta^\top x \, ,
\end{equation}
and $p + k - 1$ scalar model parameters
$\theta = (\alpha_1, \ldots, \alpha_{k-1}, \beta_1, \ldots,
\beta_p)^\top$. Straightforward algebra starting from~\eqref{eq:po}
gives that
\begin{equation}
  \label{eq:po2}
  \frac{\pi_{j}(x_2)}{\pi_{j+1}(x_2)} = \exp\{\beta^\top(x_2 - x_1)\}\frac{\pi_{j}(x_1)}{\pi_{j+1}(x_1)} \quad \text{for any } x_1, x_2 \in \Re^p \text{ and } j \in \{1, \ldots, k-1\}\, .
\end{equation}
As a result, adjacent-categories odds are indeed proportional with a
constant of proportionality that does not depend on the category. The
NPO version of the ACL model has
\begin{equation}
  \label{eq:npolp}
  \eta_{j}(x) = \alpha_j + \beta_j^\top x \, ,
\end{equation}
with $(k - 1)(p + 1)$ scalar model parameters
$\theta = (\alpha_1, \ldots, \alpha_{k-1}, \beta_1^\top, \ldots,
\beta_{k-1}^\top)^\top$, where
$\beta_j = (\beta_{j1}, \ldots, \beta_{jp})^\top$. Figure~\ref{fig:po}
shows the adjacent-categories log-odds for two distinct categories
under the PO and the NPO versions of the ACL model, for
$x \in \Re$. Note that under the PO version of the model the log-odds
for distinct categories are parallel lines, which in turn
implies~\eqref{eq:po2} for any pair of categories. On the other hand,
under the NPO version of the model the log-odds are not parallel
lines, so~\eqref{eq:po2} is not generally satisfied.

The most general version of the ACL model is the partial proportional
odds model with
\[
  \eta_{j}(x) = \alpha_j + \xi_j^\top x^{(np)} + \rho^\top x^{(p)} \,
  ,
\]
where $x^{(np)}$ and $x^{(p)}$ are sub-vectors of $x$ with distinct
components characterizing the PO and NPO effects, respectively. All
subsequent derivations, results, and discussions can be written in
terms of the more general partial proportional odds version, and then
PO and NPO can be presented as special cases. Nevertheless, we focus
on the PO and NPO versions separately, to keep the notation concise,
and because some of the following results are specific to PO and not
to NPO.

Expressions~\eqref{eq:polp} and~\eqref{eq:npolp} immediately imply that
the ACL model provides valid category probabilities across the
parameter space and regardless of whether the local odds
$\pi_{j}(x) / \pi_{j+1}(x)$ are modelled as proportional or
non-proportional. This is in contrast to other popular
ordinal-response regression models, like cumulative-logit models
\citep{mccullagh:1980}, whose NPO versions
\citep{peterson+harrell:1990} may provide invalid category
probabilities in subsets of the parameter space and covariate space,
and, hence, result in hard-to-circumvent issues with estimation,
inference, and prediction.

\subsection{Equivalence with baseline-category logit models}
\label{sec:acl_is_bcl}

Writing
$\log\{ \pi_{j}(x) / \pi_{k}(x) \} = \sum_{l = j}^{k - 1} \log\{
\pi_{l}(x) / \pi_{l+1}(x) \}$, it is simple to show that both the PO
and NPO versions of the ACL model for ordinal responses can be written
as BCL models for nominal responses
\citep[see][Section~4.1]{agresti:2010} where the $k$ category is used
as reference.

In particular, the NPO version of the ACL model in~\eqref{eq:npolp} is
equivalent to a BCL model with
\begin{equation}
  \label{eq:npo_bcl}
  \log \frac{\pi_{j}(x)}{\pi_{k}(x)} = \gamma_j + \delta_j^\top x \quad (j = 1, \ldots, k-1) \, ,
\end{equation}
where $\gamma_j = \sum_{l = j}^{k-1} \alpha_l$ and
$\delta_j = \sum_{l = j}^{k-1} \beta_l$. The PO version of the ACL
model in~\eqref{eq:polp} is equivalent to a BCL model with
\begin{equation}
  \label{eq:po_bcl}
  \log \frac{\pi_{j}(x)}{\pi_{k}(x)} = \gamma_j + (k - j) \zeta^\top x \quad (j = 1, \ldots, k-1) \, ,
\end{equation}
where $\gamma_j = \sum_{l = j}^{k-1} \alpha_l$ and $\beta =
\zeta$. 

\subsection{Maximum likelihood estimation}
\label{sec:maxim-likel-estim}

A consequence of the equivalence between the BCL and ACL models is
that we can estimate the latter using the ML
estimates for the former.  The equivariance of the maximum ML
estimator under one-to-one transformations of the model parameters
guarantees that after computing the ML estimates for the parameters of
BCL model~\eqref{eq:npo_bcl}, the model parameters of the NPO version
of the ACL can be estimated as
$\hat\alpha_j = \hat\gamma_j - \hat\gamma_{j+1}$ and
$\hat\beta_j = \hat\delta_j - \hat\delta_{j+1}$ $(j = 1, \ldots, k-1)$
with $\hat\gamma_k = 0$ and $\hat\beta_k = 0_p$, where $0_p$ is a
$p$-vector of zeros. Correspondingly, once the ML estimates for the
parameters of BCL model~\eqref{eq:po_bcl} have been obtained, the
model parameters of the PO version of the ACL model can be estimated
as $\hat\alpha_j = \hat\gamma_j - \hat\gamma_{j+1}$ and
$\hat\beta = \hat\zeta$ $(j = 1, \ldots, k-1)$.

So, ML estimation of ACL models can be performed using ready ML
implementations for fitting the BCL models~\eqref{eq:npo_bcl}
and~\eqref{eq:po_bcl}, like the \texttt{multinom()} function of the
\texttt{nnet} R package \citep{venables+ripley:2002} that exploits the
equivalence of BCL models with neural networks, and the
\texttt{brmultinom()} function of the \texttt{brglm2} R package
\citep{brglm2} that exploits the equivalence of BCL models with
Poisson log-linear models.

\subsection{Exponential families}
\label{sec:adjac-categ-logit}

The BCL model is a full exponential family distribution with natural
parameters $\gamma_j$ and $\delta_j$ for the NPO
version~(\ref{eq:npo_bcl}) of the ACL model, and $\gamma_j$ and
$\zeta$ for the PO version~(\ref{eq:po_bcl}) of the ACL model
$(j = 1, \ldots, k - 1)$. Hence, another consequence of the
equivalence of ACL models to BCL models is that both the PO and NPO
versions of the ACL model are full exponential families. Specifically,
the sufficient statistics in the NPO parameterization are
$\sum_{l = 1}^j \sum_{i = 1}^n y_{il}$ for $\alpha_j$,
$\sum_{l = 1}^j \sum_{i = 1}^n y_{il}x_{i}$ for $\beta_j$, and
$\sum_{l = 1}^j \sum_{i = 1}^n y_{il}$ for $\alpha_j$ and
$\sum_{j = 1}^{k-1} \sum_{i = 1}^n (k - j) y_{ij} x_i$ for $\beta$ in
the PO parameterization $(j = 1, \ldots, k-1)$ \citep[see, also,
][Section~4.1]{agresti:2010}.

\subsection{Infinite maximum likelihood estimates}

As is the case for their equivalent BCL models, depending on the data
configuration, the ML estimates of ACL models can have infinite
components, resulting in issues for both iterative estimation
procedures and for first-order inference about the parameters. In fact,
infinite ML estimates for the PO and NPO versions of the ACL model
result if, and only if, separation occurs for the equivalent BCL
models. Example~\ref{ex:wine} below uses a real data set to illustrate
the consequences that separation can have in the estimation of, and
inference from, ACL models.

\begin{example}
  \label{ex:wine}
  {\bf Infinite ML\ estimates in ACL models} The data set in
  Table~\ref{tab:wine_data} comes from \citet{randall:89} and concerns
  an experiment for investigating factors that affect the bitterness
  of white wine. There are two factors in the experiment, namely temperature
  at the time of crushing the grapes (with two levels, ``cold'' and
  ``warm'') and contact of the juice with the skin (with two levels
  ``Yes'' and ``No''). For each combination of factors two bottles
  were rated on their bitterness by a panel of 9 judges. The responses
  of the judges on the bitterness of the wine were taken on a
  continuous scale in the interval from 0 (``None'') to 100
  (``Intense'') and then they were grouped correspondingly into $5$
  ordered categories, labelled as ``1'', ``2'', ``3'', ``4'', and ``5''.

\begin{table}[t]
  \caption{The wine tasting data \citep{randall:89}.}
  \begin{center}
  \begin{tabular}{ccccccc}
    \toprule
    \multirow{2}{*}{Temperature} & \multirow{2}{*}{Contact} &
         \multicolumn{5}{c}{Bitterness rating} \\
          & & 1 & 2 & 3 & 4 & 5 \\ \midrule
    Cold & No  & 4 & 9 & 5 & 0 & 0 \\
    Cold & Yes & 1 & 7 & 8 & 2 & 0 \\
    Warm & No  & 0 & 5 & 8 & 3 & 2 \\
    Warm & Yes & 0 & 1 & 5 & 7 & 5 \\ \bottomrule
  \end{tabular}
\end{center}
\label{tab:wine_data}
\end{table}

\begin{figure}
  \begin{center}
  \includegraphics[scale=0.8]{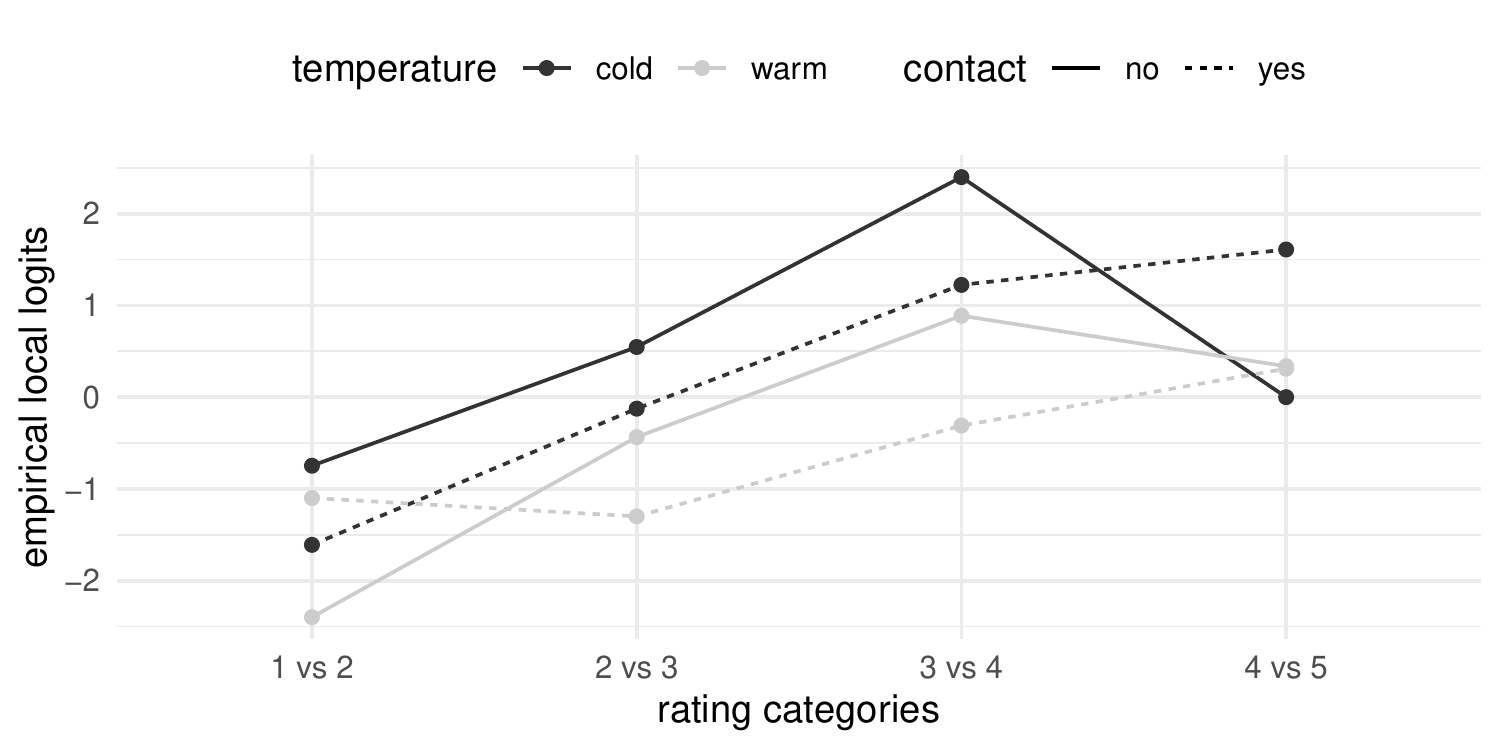}  
\end{center}
\caption{The empirical adjacent logits
  $\log\{(y_{ij} + 1/2)/(y_{ij+1} + 1/2)\}$ $(j = 1, \ldots, 4)$ for
  the bitterness rating for all combinations of levels for temperature
  and contact.}
\label{fig:wine_elogits}
\end{figure}

Figure~\ref{fig:wine_elogits} shows the empirical adjacent logits
$\log\{(y_{ij} + 1/2)/(y_{ij+1} + 1/2)\}$ $(j = 1, \ldots, 4)$ for the
bitterness rating for all combinations of temperature and contact.
Note that $1/2$ has been added to all frequencies as a means of
getting estimates of the adjacent-categories logits with second-order
mean bias \citep[see, for example,][]{haldane:1955}, avoiding infinite
estimates in the process.

There seems to be evidence that the adjacent logits for the
combinations of temperature and contact are parallel (see, also,
Figure~\ref{fig:po}), or in other words, the adjacent odds ratios
across temperature and/or contact levels do not depend on the
rating. The latter hypothesis can be formally tested by estimating the
NPO version of the ACL model
\begin{equation}
  \label{eq:wine_npo}
  \log \frac{\pi_{j}(t, c)}{\pi_{j + 1}(t, c)} = \alpha_j + \beta_{1j} t + \beta_{2j} c \quad (j = 1, \ldots, 4) \,,
\end{equation}
where $t$ is $1$ if temperature is warm and $0$ otherwise, $c$ is $1$
if contact is yes and $0$ otherwise, and $\pi_{j}(t, c)$ is the
probability of a bitterness rating $j$ at $t$ and $c$. The hypotheses
of parallel adjacent logits can then be written in terms of the model
parameters as $\beta_{11} = \ldots = \beta_{14} = \beta_1$ and
$\beta_{21} = \ldots = \beta_{24} = \beta_2$, and tested using the
value of the Wald statistic
\begin{equation}
  \label{eq:wald_stat}
  W = \hat{\theta}^\top C^\top \left\{C i(\hat{\theta})^{-1} C^\top\right\}^{-1} C \hat{\theta} \, ,
\end{equation}
where $\hat{\theta}$ is the ML estimate of
$\theta = (\alpha_1, \ldots, \alpha_4, \beta_{11}, \ldots, \beta_{14},
\beta_{21}, \ldots, \beta_{24})^\top$ for model~(\ref{eq:wine_npo}),
and $i(\theta)$ is the expected information matrix at $\theta$. The
contrast matrix $C$ we use in~(\ref{eq:wald_stat}) has the form
\[
  C = \left[
    \begin{array}{ccc}
      0_{3 \times 4} & C_1 & 0_{3 \times 4} \\
      0_{3 \times 4} & 0_{3 \times 4} & C_1
    \end{array}
  \right] \quad \text{with} \quad
  C_1 = \left[
    \begin{array}{cccc}
      1 & 0 & 0 & -1 \\
      0 & 1 & 0 & -1 \\
      0 & 0 & 1 & -1
    \end{array}
  \right]\, ,
\]
where $0_{a \times b}$ is an $a \times b$ matrix of zeros. General
results about the limiting distribution of the ML estimator under mild
regularity conditions (see, for example, \citealp[Section~7.1 and
Section 7.2]{mccullagh:2018} and
\citealp[Section~9.1]{cox+hinkley:1974}) can be used to show that the
Wald statistic has asymptotically a $\chi^2_6$ distribution.

\begin{table}[t]
  \caption{Top: ML estimates and estimated standard errors (in
    parenthesis) from fitting the ACL model in~\eqref{eq:wine_npo} on
    the data in Table~\ref{tab:wine_data}. The estimates are obtained
    using the \texttt{vglm()} function of the \texttt{VGAM} R package
    \citep{VGAM} version 1.1-5 with default converge criteria
    (\texttt{epsilon = $10^{-7}$} in \texttt{vglm.control()}). Bottom:
    ML estimates and estimated standard errors using stricter
    convergence criteria (\texttt{epsilon = $10^{-9}$ in
      \texttt{vglm.control()}}). The estimated standard errors are
    computed as the square roots of the diagonal of the inverse of the
    expected information matrix at the ML estimates. The column
    $\ell(\hat\theta)$ gives the maximized log-likelihood for each
    fit.}
  \begin{center}
    \begin{tabular}{
      c@{\extracolsep{0.6cm}}
      c@{\extracolsep{0.6cm}}
      c@{\extracolsep{0.6cm}}
      r@{\extracolsep{0.2cm}}
      r@{\extracolsep{0.6cm}}
      r@{\extracolsep{0.2cm}}
      r@{\extracolsep{0.6cm}}
      r@{\extracolsep{0.2cm}}
      r}
      \toprule
  \texttt{epsilon} & $\ell(\hat\theta)$ & rating ($j$)  & \multicolumn{2}{c}{$\hat\alpha_j$} & \multicolumn{2}{c}{$\hat\beta_{1j}$} & \multicolumn{2}{c}{$\hat\beta_{2j}$} \\ 
      \midrule
 \multirow{4}{*}{$10^{-7}$} &  \multirow{4}{*}{-15.29} &  1 &  -0.83 &     (0.59) & -20.26 & (10047.96) & -1.10 & (1.21) \\
 & &  2 &   0.67 &     (0.52) &  -1.21 &     (0.66) & -0.87 & (0.64) \\
 & & 3 &   3.08 &     (1.05) &  -1.98 &     (0.92) & -1.54 & (0.83) \\
 & & 4 &   20.22 & (10732.18) & -19.89 & (10732.18) &  0.04 & (1.08) \\
      \midrule
 \multirow{4}{*}{$10^{-9}$} &  \multirow{4}{*}{-15.29} &  1 &  -0.83 &     (0.59) & -25.26 & (122409.18) & -1.10 & (1.21) \\
 & & 2 &   0.67 &     (0.52) &  -1.21 &     (0.66) & -0.87 & (0.64) \\
 & & 3 &   3.08 &     (1.05) &  -1.98 &     (0.92) & -1.54 & (0.83) \\
 & & 4 &   25.22 & (130748.2) & -24.89 & (130748.24) &  0.04 & (1.08) \\
      \bottomrule
    \end{tabular}
  \end{center}
  \label{tab:npo_ml}
\end{table}

Table~\ref{tab:npo_ml} shows the ML estimates of the ACL model
in~\eqref{eq:wine_npo}, as computed using the \texttt{vglm()} function
of the \texttt{VGAM} R package \citep{VGAM}. No warnings or
errors were returned when fitting the model. As has been the case in
the logistic regression model of Example~\ref{ex:separation}, the
estimates and estimated standard errors for $\alpha_4$, $\beta_{11}$
and $\beta_{14}$ are atypically large in absolute value. It is also
clear that these estimates and estimated standard errors increase in
absolute value as the convergence criteria get stricter, while the
maximized log-likelihood value remains the same to the displayed
accuracy.

These issues are not due to the implementation of the \texttt{vglm()}
function; instead they are consequences of quasi-complete separation
for this particular combination of data and
model~\eqref{eq:wine_npo}. The ML estimates $\hat{\alpha}_4$,
$\hat{\beta}_{11}$ and $\hat{\beta}_{14}$ in Table~\ref{tab:npo_ml}
are formally $\infty$, $-\infty$ and $-\infty$, the corresponding
estimated standard errors are all $\infty$, and the likelihood surface
has an asymptote at $-15.29$ as $\alpha_4$, $\beta_{11}$ and
$\beta_{14}$ diverge to $\infty$, $-\infty$ and $-\infty$,
respectively, along a ray in the parameter space.

Note here that the estimated standard errors appear to diverge faster
than the ML estimates do as the convergence criteria get stricter. As
a result, the typically reported $Z$-statistics for individual
hypothesis tests about the parameters will tend to be spuriously small
in absolute value regardless of the strength of the evidence against
the hypotheses. Hence, the naive use of the computer output for
inference about the parameters of ACL models is likely to lead to
invalid conclusions when data separation occurs. More importantly,
having estimates on the boundary of the parameter space violates the
assumptions required for the asymptotic $\chi^2$ distribution
of~\eqref{eq:wald_stat}. Consequently, it is hard to justify the performance and
validity of the Wald statistic in that case.

\end{example}

\section{Mean and median bias reduction for ACL models}
\label{sec:acl_meanBR}

A consequence of the ACL models being full exponential family
distributions (see Section~\ref{sec:adjac-categ-logit}) is that mean
BR can be implemented by maximizing the penalized likelihood
in~(\ref{eq:penloglik}). Nevertheless, as for ML, mean BR estimates
for ACL models can be conveniently computed through a ready
implementation for mean BR in BCL models coupled with the equivariance
of the mean BR estimator under linear transformations (see
Section~\ref{sec:meanBR_trans}).

\citet{kosmidis+firth:2011} prove that the equivalence of BCL models
and Poisson log-linear models (see, also, \citealp{palmgren:1981} and
\citealp{baker:1994} for authoritative descriptions of that
equivalence) extends to the mBR estimates, and describe a simple
algorithm for mBR estimation of BCL models, each iteration of which
consists of the following steps:
\begin{itemize}
\item[P1] Rescale the Poisson means to match the observed multinomial totals.
\item[P2] Add half a leverage based on the rescaled means to the
  observed multinomial frequencies.
\item[P3] Estimate, using ML, the equivalent Poisson log-linear model to
  the adjusted frequencies.
\end{itemize}
Iteration stops when the differences between successive estimates or,
alternatively, the mean BR adjusted scores in~(\ref{eq:meanbr}) are
smaller than a pre-determined, small positive constant. An alternative
criterion can be based on the change of the mean BR penalized
likelihood~(\ref{eq:penloglik}) between successive iterations.  mBR
estimates for ACL models can then be computed as follows
\begin{itemize}
\item[S1] Compute mBR estimates of the parameters $\gamma_j$ and
  $\delta_j$ of the BCL model in~\eqref{eq:npo_bcl} for the NPO
  version (or $\gamma_j$ and $\zeta$ of the BCL model
  in~\eqref{eq:po_bcl} for the PO version) $(j = 1, \ldots, q)$ by
  iterating steps P1, P2, and P3.
\item[S2] Calculate the mBR estimates for the NPO version of the ACL
  model as $\alpha_j^* = \gamma_j^* - \gamma_{j+1}^*$ and
  $\beta_j^* = \delta_j^* - \delta_{j+1}^*$ (or
  $\alpha_j^* = \gamma_j^* - \gamma_{j+1}^*$ and $\beta^* = \zeta^*$
  for the PO version) $(j = 1, \ldots, q)$, with $\gamma_k^* = 0$ and
  $\beta_k^* = 0_p$.
\end{itemize}

Implementation of mdBR for ACL models is not as direct as that of
mBR. A maximum penalized likelihood interpretation of mdBR does not
exist for general ACL models, like it does for mBR. Also, since
contrasts of parameters are not component-wise transformations,
algorithms for mdBR for BCL models (see
\citealp[Section~6]{kosmidis+kennepagui+sartori:2020} for extensions
of the results in \citealp{kosmidis+firth:2011}) can only be used to
get mdBR estimates $\beta^\dagger$ of $\beta$ in the PO version of the
ACL model. In other words, the estimates
$\gamma_j^\dagger - \gamma_{j+1}^\dagger$ and
$\beta_j^\dagger = \delta_j^\dagger - \delta_{j+1}^\dagger$
$(j = 1, \ldots, q)$ are not mdBR estimates, unless $k = 2$. Hence,
for general ACL models, computing the mdBR estimates $\theta^\dagger$
must rely on implementing and solving the mdBR adjusted score
equations~(\ref{eq:medianbr}). That can certainly be done (using, for
example, the quasi-Fisher scoring iteration~(\ref{eq:quasi_fisher})),
with the only effort being in deriving $P_t(\theta)$ using the
expressions for mBR in BCL models in
\citet[Appendix~B.5]{kosmidis:2007}.

\begin{example}
  \label{ex:wine2}
  {\bf Infinite ML estimates in ACL models (continued)}
  Table~\ref{tab:npo_meanbr} gives the mBR estimates from fitting the
  ACL model in~\eqref{eq:wine_npo} on the data in
  Table~\ref{tab:wine_data}. The mBR estimates are computed using the
  $\texttt{bracl()}$ function of the \texttt{brglm2} R package, which
  implements mBR through the corresponding Poisson log-linear model,
  as detailed earlier.  No convergence issues have been reported; the
  absolute values of the components of the adjusted score functions
  in~\eqref{eq:meanbr} at the mBR estimates are all less than
  $10^{-6}$, and all estimates and estimated standard errors remain
  unchanged to the reported accuracy as the convergence criteria get
  stricter.

  The Wald statistic~\eqref{eq:wald_stat} when $\hat\theta$ is
  replaced by $\theta^*$ has value $1.067$, which is small compared to
  the value of the $95\%$ quantile of a $\chi^2_6$ distribution
  ($12.592$), providing no evidence against the simpler PO model with
  $\beta_{11} = \ldots = \beta_{14} = \beta_1$ and
  $\beta_{21} = \ldots = \beta_{24} = \beta_2$.
 
  \begin{table}[t]
    \caption{Mean BR estimates and estimated standard errors (in
      parenthesis) from fitting the ACL model in~\eqref{eq:wine_npo}
      on the data in Table~\ref{tab:wine_data}. The estimates are
      obtained using the \texttt{bracl()} function of the
      \texttt{brglm2} R package \citep{brglm2} version 0.7.2 with
      default convergence criteria. The estimated standard errors are
      computed as the square roots of the diagonal of
      $i(\theta^*)^{-1}$.}
  \begin{center}
    \begin{tabular}{
      c@{\extracolsep{0.6cm}}
      r@{\extracolsep{0.2cm}}
      r@{\extracolsep{0.6cm}}
      r@{\extracolsep{0.2cm}}
      r@{\extracolsep{0.6cm}}
      r@{\extracolsep{0.2cm}}
      r}
      \toprule
      rating ($j$)  & \multicolumn{2}{c}{$\alpha_j^*$} & \multicolumn{2}{c}{$\beta_{1j}^*$} & \multicolumn{2}{c}{$\beta_{2j}^*$} \\ 
      \midrule
      1 & -0.76  & (0.59) & -1.65  & (1.60) & -0.82  & (1.08) \\
      2 & 0.62  & (0.52) & -1.12  & (0.66) & -0.80  & (0.64) \\  
      3 & 2.73  & (0.99) & -1.75  & (0.87) & -1.38  & (0.81) \\ 
      4 & 1.53  & (1.83) & -1.26  & (1.68) & 0.07  & (1.03) \\ 
      \bottomrule
    \end{tabular}
  \end{center}
  \label{tab:npo_meanbr}
\end{table}

Comparing the mBR estimates in Table~\ref{tab:npo_meanbr} to the ML
ones in Table~\ref{tab:npo_ml}, we notice that the mBR estimates are
shrunken towards zero relative to ML ones. As a result, the fitted
multinomial probabilities at the mBR estimates are closer to
$(1/5, 1/5, 1/5, 1/5, 1/5)^\top$ than ones at the ML estimates. In
other words, mBR shrinks the model towards equi-probability across
observations.  This is a generalization of the shrinkage effect of mBR
we observed in Example~\ref{ex:separation2} and that
\citet{kosmidis+firth:2021} study theoretically in the special case of
logistic regression ($k = 2$).

It is interesting to note that the shrinkage direction of mBR in
cumulative logit models for global cumulative odds
\citep{kosmidis:2014} is rather different; the fitted multinomial
probabilities at the mBR estimates for the PO version of the
cumulative logit model would be closer to $(1/2, 0, 0, 0, 1/2)^\top$
than the ones at the ML estimates. In other words, mBR shrinks the
cumulative logit model towards a logistic regression model for the end
categories.

The ML, mdBR, and mBR estimates for $\beta_1$ for the PO version of the
ACL model are $-1.69$, $-1.61$, and $-1.56$, respectively, with
corresponding estimated standard errors $0.41$, $0.39$, and
$0.38$. The respective estimates for $\beta_2$ are $-0.96$, $-0.92$,
and $-0.90$, respectively, with corresponding estimated standard
errors $0.32$, $0.31$, and $0.31$. The shrinkage towards
equi-probability that mBR delivers is also apparent in the estimates
for the PO version of the ACL model. As is the case in logistic
regression, mdBR also tends to shrink estimates towards zero, but that
shrinkage effect is less strong than from mBR.

\end{example}

\section{Mean bias reduction of ordinal superiority summaries}
\label{sec:meanBR_superiority}

The mean BR estimates for ACL models can be used to get improved
estimates of other model summaries by using the bias of
transformations of the mean RB estimator in
expression~(\ref{eq:bias_trans}).

A prominent example of such a summary are the ordinal, model-based
superiority measures for comparing distributions of two groups,
adjusted for covariates that are introduced in
\citet{agresti+kateri:2017}.  In ordinal-response models with a latent
variable interpretation, such as cumulative-link models
\citep{mccullagh:1980}, ordinal superiority measures can be defined
directly on the latent scale, which results in exact (for probit,
log-log, and complementary log-log link) or approximate expressions
(for logit link) that are functions of only the coefficient of the
indicator variable characterizing the two groups being compared. This
fact has been exploited in \citet{gioia+kennepagui+salvan:2021}, who
used the equivariance properties of the mdBR estimator (see
Section~\ref{sec:medianBR_trans}) to directly transform the mdBR
estimates of the group indicator parameter to deliver mdBR estimates
of ordinal superiority measures.

In more general models for ordinal responses that may also lack a
latent variable interpretation (like ACL models), ordinal superiority
measures are instead defined in terms of category probabilities that
necessarily depend on all model parameters. Suppose that the covariate
vector is $(w^\top, z)^\top$, where $z$ is a group indicator variable
taking value $0$ for group $1$ and value $0$ for group $2$, and denote
by $\pi_{j}(w, 1)$ and $\pi_{j}(w, 0)$ $(j =1, \ldots, k)$ the
model-based probabilities of category $j$ at covariate values $w$, for
group 1 and group 2, respectively. The dependence of the probabilities
on the model parameters has been suppressed here for notational
convenience.

\citet{agresti+kateri:2017} propose comparing the distribution of the
ordinal response at group 1 to that at group 2, at covariate values
$w$, through the ordinal superiority measure
\begin{equation}
  \label{eq:delta}
  \Delta(w; \theta) = \sum_{r > s} \pi_{r}(w, 1) \pi_{s}(w, 0) - \sum_{s > r} \pi_{r}(w, 1) \pi_{s}(w, 0) \, .
\end{equation}
If the two distributions are identical then $\Delta(w; \theta) =
0$. Positive values of $\Delta(w; \theta)$ indicate that for
covariates $w$, it is more likely to observe higher response
categories in group 1 than in group 2, and vice versa for negative
values. A related ordinal superiority measure is
\begin{equation}
  \label{eq:gamma}
  \gamma(w; \theta) = 2 \Delta(w; \theta) - 1 \, ,
\end{equation}
which takes values between $0$ and $1$, and is interpreted as the
probability that the response category in group 1 is higher than the
response category in group 2, while adjusting for covariates $w$
\citep[see][for details]{klotz:1966}. In practice, the covariate
setting $w$ can be taken to be a representative value from a sample of
covariate values $w_1, \ldots, w_n$, e.g.
$\bar{w} = \sum_{i = 1}^n w_i /n$. Alternatively, if the sample of
covariate values is representative of the population of interest then
summary ordinal superiority measures can be defined as
\begin{equation}
  \label{eq:summaryosm}
  \bar{\Delta}(\theta) = \frac{1}{n} \sum \Delta(w_i; \theta) \quad \text{and} \quad \bar\gamma(\theta) = \frac{1}{n} \sum \gamma(w_i; \theta) \, .
\end{equation}

\citet{agresti+kateri:2017} propose estimating the ordinal
superiority measures by replacing $\theta$ in
expressions~\eqref{eq:delta}, \eqref{eq:gamma},
and~\eqref{eq:summaryosm} by the ML estimator $\hat\theta$, and use
the delta method to construct inferences about those measures. Note
here that because of the specific equivariance properties of the mBR
and mdBR estimator (see Section~\ref{sec:bias_trans}), replacing
$\theta$ by the mBR estimator $\theta^*$ or a mdBR estimator
$\theta^\dagger$ does not, in general, result in mBR or mdBR estimators
of the measures. In fact, despite it being the case that the resulting estimators will
be consistent under the same conditions that their ML counterparts
are, they may end up having much worse finite-sample mean and/or
median bias properties than the ML version does.

mdBR estimators of \eqref{eq:delta}, \eqref{eq:gamma},
and~\eqref{eq:summaryosm} are not easy to construct. In contrast, an
easy-to-compute mBR estimator of $\Delta(w; \theta)$ and of the other
ordinary superiority measures can be derived using
expression~(\ref{eq:bias_trans}). In particular, an mBR estimator of
$\Delta(w; \theta)$ is
\[
  \Delta^*(w; \theta^*) = \Delta(w; \theta^*) - B^*(w; \theta^*) \, .
\]
where
\[
  B^*(w; \theta) = \frac{1}{2} \trace\left\{i(\theta)^{-1}
    \nabla\nabla^\top \Delta(w; \theta)\right\} \, ,
\]
is the first term in the right-hand side of
expression~(\ref{eq:bias_trans}). Computing $\Delta^*(w; \theta^*)$
requires only the mBR estimator $\theta^*$ that can be obtained using
the procedures in Section~\ref{sec:acl_meanBR}, the corresponding
estimated category probabilities at $(w^\top, 1)$ and $(w^\top, 0)$,
the matrix $i(\theta^*)^{-1}$, and the hessian
$\nabla \nabla^\top \Delta(w; \theta^*)$. All these quantities, except
$\nabla \nabla^\top \Delta(w; \theta^*)$, are readily available or can
be readily computed once the model has been estimated using mBR, as is
done, for example, in Section~\ref{sec:meanBR} for ACL models and in
\citet{kosmidis:2014} for cumulative link models. For specific
ordinal-response models, the hessian
$\nabla \nabla^\top \Delta(w; \theta)$ can be analytically obtained
with some algebraic effort. For example, if $\pi_{j}(w, z)$ is based
on cumulative link models one can work with the expressions for the
derivatives of $\Delta(w; \theta)$ in \citet[Web appendix
A]{agresti+kateri:2017}. Alternatively, a very accurate approximation
of $\nabla \nabla^\top \Delta(w; \theta^*)$ can be obtained for
general models using a ready implementation of $\Delta(w; \theta)$ and
numerical differentiation routines, like the ones provided in the
\texttt{numDeriv} R package \citep{numDeriv}. This is the route that
the \texttt{ordinal\_superiority()} method of the \texttt{brglm2} R
package takes.

Due to the equivariance properties of mBR estimation in
Section~\ref{sec:meanBR_trans} under linear transformations, mBR
estimators of $\gamma(w; \theta)$, $\Delta^\dagger(\theta)$, and
$\gamma^\dagger(\theta)$ are readily obtained by replacing
$\Delta(w; \theta)$ by $\Delta^*(w; \theta^*)$ in
expressions~\eqref{eq:gamma} and~\eqref{eq:summaryosm}. Wald-type
inferences about the mBR estimators of the ordinal superiority
measures can be constructed as proposed in
\citet[Section~5]{agresti+kateri:2017}, using the mBR estimates of the
ordinal superiority measures along with estimated standard errors
obtained using the delta method, based on $i(\theta^*)$, and numerical
gradients.

\begin{example}
  {\bf mBR for ordinal superiority measures from ACL model} In order
  to assess the finite sample properties of the mBR estimator of
  ordinal superiority scores in ACL models, we consider the example in
  \citet[Section~4.3]{ordinal}, where the bitterness ratings
  ``2'', ``3'', and ``4'' in Table~\ref{tab:wine_data} are merged into a
  single rating ``2-4''. Like the PO version of the cumulative logit
  model \citep[see ][Section~4.8]{ordinal}, the ML estimates
  for $\alpha_{2-4}$ and $\beta_1$ for the PO version of the ACL
  model~\eqref{eq:wine_npo} are $+\infty$ and $-\infty$
  respectively. The mBR estimates of
  $\theta = (\alpha_1, \alpha_{2-4}, \beta_1, \beta_2)^\top$, on the
  other hand, take the finite values $\theta^* = (-1.247$, $5.331$,
  $-3.291$, $-1.181)^\top$.  If $\gamma(w, \theta)$ is the ordinal
  superiority measure for temperature setting $w$ ($w = 0$ for cold
  and $w = 1$ for warm), and $z$ indicates contact ($z = 1$) or not
  ($z = 0$) of the juice with the skin, then
  $\gamma(0, \theta^*) = 0.594$ and $\gamma(1, \theta^*) = 0.575$,
  indicating that there is almost $60\%$ chance of higher bitterness
  ratings when there is contact of the juice with the skin.

  We simulate $10,000$ samples from the PO version of the ACL model
  at $\bar\theta$, and we compute $\gamma(w, \hat\theta)$ and
  $\gamma^*(w, \theta^*)$ for each sample. The simulation-based
  estimates of the finite-sample relative biases of
  $\gamma(w, \hat\theta)$ are $0.84\%$ and $1.56\%$ for $w = 0$ and
  $w = 1$, respectively. As expected, the mBR version
  $\gamma^*(w, \theta^*)$ is found to have smaller finite-sample
  relative biases at $0.13\%$ and $-0.02\%$ for $w = 0$ and $w = 1$,
  respectively. The corresponding percentages of underestimation are
  $48.48\%$ and $44.69\%$ for $\gamma(w, \hat\theta)$, and $52.12\%$
  and $51.14\%$ for $\gamma^*(w, \theta^*)$. Hence, in this case, mBR
  also results in improvements in median bias. Finally, both
  estimators appear to perform satisfactorily in terms of Wald-type
  inferences based on them. The coverage probability of the nominally
  $95\%$ Wald-type confidence intervals based on
  $\gamma(w, \hat\theta)$ are $94.8\%$ ($w = 0$) and $94.6\%$
  ($w = 1$), and $94.7\%$ ($w = 0$) and $95.1\%$ ($w = 1$) for those
  based on $\gamma^*(w, \theta^*)$.
\end{example}

\section{Supplementary material}
The supplementary material consists of three scripts that replicate
all the numerical results and graphics reported in the paper, and is
available at
\url{https://ikosmidis.com/files/bracl_supplementary_v0.2.zip}. The
results are exactly reproducible in R version 4.1.2, and with the
following packages: \texttt{VGAM} version 1.1-5 \citep{VGAM},
\texttt{tibble} 3.1.6 \citep{tibble}, \texttt{dplyr} 1.0.7
\citep{dplyr}, \texttt{ggplot2} 3.3.5 \citep{ggplot2},
\texttt{colorspace} 2.0-2 \citep{colorspace}, and \texttt{ordinal}
2019.12-10 \citep{ordinal}, \texttt{brglm2} 0.8.2 \citep{brglm2},
\texttt{enrichwith} 0.3.1 \citep{enrichwith}, and
\texttt{detectseparation} 0.2 \citep{detectseparation}.

\section{Acknowledgements}

The author greatly appreciates the constructive discussions with Alan
Agresti and Anestis Touloumis during the Challenges for Categorical
Data Analysis 2018 Workshop in Aachen University on the equivalence
between ACL and BCL models, which informed this work. Ioannis Kosmidis
has been partially supported by The Alan Turing Institute under the
EPSRC grant EP/N510129/1.

\bibliographystyle{chicago}
\bibliography{bracl}

\end{document}

%% file: figures/PO.tex
\ifx\du\undefined
  \newlength{\du}
\fi
\setlength{\du}{12\unitlength}
\begin{tikzpicture}
\pgftransformxscale{1.000000}
\pgftransformyscale{-1.000000}
\definecolor{dialinecolor}{rgb}{0.000000, 0.000000, 0.000000}
\pgfsetstrokecolor{dialinecolor}
\definecolor{dialinecolor}{rgb}{1.000000, 1.000000, 1.000000}
\pgfsetfillcolor{dialinecolor}
\pgfsetlinewidth{0.050000\du}
\pgfsetdash{{\pgflinewidth}{0.200000\du}}{0cm}
\pgfsetdash{{\pgflinewidth}{0.200000\du}}{0cm}
\pgfsetbuttcap
{
\definecolor{dialinecolor}{rgb}{0.000000, 0.000000, 0.000000}
\pgfsetfillcolor{dialinecolor}
\definecolor{dialinecolor}{rgb}{0.000000, 0.000000, 0.000000}
\pgfsetstrokecolor{dialinecolor}
\draw (38.775145\du,27.700000\du)--(55.975145\du,27.600000\du);
}
\pgfsetlinewidth{0.050000\du}
\pgfsetdash{}{0pt}
\pgfsetdash{}{0pt}
\pgfsetbuttcap
{
\definecolor{dialinecolor}{rgb}{0.000000, 0.000000, 0.000000}
\pgfsetfillcolor{dialinecolor}
\definecolor{dialinecolor}{rgb}{0.000000, 0.000000, 0.000000}
\pgfsetstrokecolor{dialinecolor}
\draw (45.983750\du,15.740000\du)--(45.983750\du,15.740000\du);
}
\pgfsetlinewidth{0.050000\du}
\pgfsetdash{{\pgflinewidth}{0.200000\du}}{0cm}
\pgfsetdash{{\pgflinewidth}{0.200000\du}}{0cm}
\pgfsetbuttcap
{
\definecolor{dialinecolor}{rgb}{0.000000, 0.000000, 0.000000}
\pgfsetfillcolor{dialinecolor}
\definecolor{dialinecolor}{rgb}{0.000000, 0.000000, 0.000000}
\pgfsetstrokecolor{dialinecolor}
\draw (40.625145\du,27.550000\du)--(40.725145\du,12.200000\du);
}
\pgfsetlinewidth{0.050000\du}
\pgfsetdash{{\pgflinewidth}{0.200000\du}}{0cm}
\pgfsetdash{{\pgflinewidth}{0.200000\du}}{0cm}
\pgfsetbuttcap
{
\definecolor{dialinecolor}{rgb}{0.000000, 0.000000, 0.000000}
\pgfsetfillcolor{dialinecolor}
\definecolor{dialinecolor}{rgb}{0.000000, 0.000000, 0.000000}
\pgfsetstrokecolor{dialinecolor}
\draw (49.550273\du,27.510128\du)--(49.675145\du,12.200000\du);
}
\definecolor{dialinecolor}{rgb}{0.000000, 0.000000, 0.000000}
\pgfsetstrokecolor{dialinecolor}
\node[anchor=west] at (40.425145\du,28.335000\du){$x_1$};
\definecolor{dialinecolor}{rgb}{0.000000, 0.000000, 0.000000}
\pgfsetstrokecolor{dialinecolor}
\node[anchor=west] at (49.275145\du,28.335000\du){$x_2$};
\definecolor{dialinecolor}{rgb}{0.000000, 0.000000, 0.000000}
\pgfsetstrokecolor{dialinecolor}
\node[anchor=west] at (53.275145\du,12.000000\du){$\log\frac{\pi_j(x)}{\pi_{j + 1}(x)}$};
\pgfsetlinewidth{0.050000\du}
\pgfsetdash{{1.000000\du}{1.000000\du}}{0\du}
\pgfsetdash{{0.300000\du}{0.300000\du}}{0\du}
\pgfsetbuttcap
{
\definecolor{dialinecolor}{rgb}{0.000000, 0.000000, 0.000000}
\pgfsetfillcolor{dialinecolor}
\definecolor{dialinecolor}{rgb}{0.000000, 0.000000, 0.000000}
\pgfsetstrokecolor{dialinecolor}
\draw (40.572820\du,20.900000\du)--(49.572820\du,20.900000\du);
}
\pgfsetlinewidth{0.050000\du}
\pgfsetdash{{0.300000\du}{0.300000\du}}{0\du}
\pgfsetdash{{0.300000\du}{0.300000\du}}{0\du}
\pgfsetbuttcap
{
\definecolor{dialinecolor}{rgb}{0.000000, 0.000000, 0.000000}
\pgfsetfillcolor{dialinecolor}
\definecolor{dialinecolor}{rgb}{0.000000, 0.000000, 0.000000}
\pgfsetstrokecolor{dialinecolor}
\draw (49.625145\du,15.350000\du)--(49.572820\du,20.950000\du);
}
\definecolor{dialinecolor}{rgb}{0.000000, 0.000000, 0.000000}
\pgfsetstrokecolor{dialinecolor}
\node[anchor=west] at (49.70000\du,17.700000\du){$\beta_j (x_2 - x_1)$};
\pgfsetlinewidth{0.050000\du}
\pgfsetdash{}{0pt}
\pgfsetdash{}{0pt}
\pgfsetbuttcap
{
\definecolor{dialinecolor}{rgb}{0.000000, 0.000000, 0.000000}
\pgfsetfillcolor{dialinecolor}
\definecolor{dialinecolor}{rgb}{0.000000, 0.000000, 0.000000}
\pgfsetstrokecolor{dialinecolor}
\draw (39.458230\du,21.438116\du)--(53.775145\du,12.900000\du);
}
\definecolor{dialinecolor}{rgb}{0.000000, 0.000000, 0.000000}
\pgfsetstrokecolor{dialinecolor}
\node[anchor=west] at (53.525145\du,20.000000\du){$\log\frac{\pi_k(x)}{\pi_{k + 1}(x)}$};
\pgfsetlinewidth{0.050000\du}
\pgfsetdash{{1.000000\du}{1.000000\du}}{0\du}
\pgfsetdash{{0.300000\du}{0.300000\du}}{0\du}
\pgfsetbuttcap
{
\definecolor{dialinecolor}{rgb}{0.000000, 0.000000, 0.000000}
\pgfsetfillcolor{dialinecolor}
\definecolor{dialinecolor}{rgb}{0.000000, 0.000000, 0.000000}
\pgfsetstrokecolor{dialinecolor}
\draw (40.622851\du,25.070000\du)--(49.622851\du,25.070000\du);
}
\pgfsetlinewidth{0.050000\du}
\pgfsetdash{{0.300000\du}{0.300000\du}}{0\du}
\pgfsetdash{{0.300000\du}{0.300000\du}}{0\du}
\pgfsetbuttcap
{
\definecolor{dialinecolor}{rgb}{0.000000, 0.000000, 0.000000}
\pgfsetfillcolor{dialinecolor}
\definecolor{dialinecolor}{rgb}{0.000000, 0.000000, 0.000000}
\pgfsetstrokecolor{dialinecolor}
\draw (49.575145\du,22.150000\du)--(49.622851\du,25.020000\du);
}
\definecolor{dialinecolor}{rgb}{0.000000, 0.000000, 0.000000}
\pgfsetstrokecolor{dialinecolor}
\node[anchor=west] at (49.70000\du,23.870000\du){$\beta_k (x_2 - x_1)$};
\pgfsetlinewidth{0.050000\du}
\pgfsetdash{}{0pt}
\pgfsetdash{}{0pt}
\pgfsetbuttcap
{
\definecolor{dialinecolor}{rgb}{0.000000, 0.000000, 0.000000}
\pgfsetfillcolor{dialinecolor}
\definecolor{dialinecolor}{rgb}{0.000000, 0.000000, 0.000000}
\pgfsetstrokecolor{dialinecolor}
\draw (39.508261\du,25.458116\du)--(53.975145\du,20.750000\du);
}
\pgfsetlinewidth{0.050000\du}
\pgfsetdash{{\pgflinewidth}{0.200000\du}}{0cm}
\pgfsetdash{{\pgflinewidth}{0.200000\du}}{0cm}
\pgfsetbuttcap
{
\definecolor{dialinecolor}{rgb}{0.000000, 0.000000, 0.000000}
\pgfsetfillcolor{dialinecolor}
\definecolor{dialinecolor}{rgb}{0.000000, 0.000000, 0.000000}
\pgfsetstrokecolor{dialinecolor}
\draw (18.450000\du,27.700000\du)--(35.650000\du,27.600000\du);
}
\pgfsetlinewidth{0.050000\du}
\pgfsetdash{}{0pt}
\pgfsetdash{}{0pt}
\pgfsetbuttcap
{
\definecolor{dialinecolor}{rgb}{0.000000, 0.000000, 0.000000}
\pgfsetfillcolor{dialinecolor}
\definecolor{dialinecolor}{rgb}{0.000000, 0.000000, 0.000000}
\pgfsetstrokecolor{dialinecolor}
\draw (25.658600\du,15.740000\du)--(25.658600\du,15.740000\du);
}
\pgfsetlinewidth{0.050000\du}
\pgfsetdash{{\pgflinewidth}{0.200000\du}}{0cm}
\pgfsetdash{{\pgflinewidth}{0.200000\du}}{0cm}
\pgfsetbuttcap
{
\definecolor{dialinecolor}{rgb}{0.000000, 0.000000, 0.000000}
\pgfsetfillcolor{dialinecolor}
\definecolor{dialinecolor}{rgb}{0.000000, 0.000000, 0.000000}
\pgfsetstrokecolor{dialinecolor}
\draw (20.300000\du,27.550000\du)--(20.400000\du,12.200000\du);
}
\pgfsetlinewidth{0.050000\du}
\pgfsetdash{{\pgflinewidth}{0.200000\du}}{0cm}
\pgfsetdash{{\pgflinewidth}{0.200000\du}}{0cm}
\pgfsetbuttcap
{
\definecolor{dialinecolor}{rgb}{0.000000, 0.000000, 0.000000}
\pgfsetfillcolor{dialinecolor}
\definecolor{dialinecolor}{rgb}{0.000000, 0.000000, 0.000000}
\pgfsetstrokecolor{dialinecolor}
\draw (29.225100\du,27.510100\du)--(29.350000\du,12.200000\du);
}
\definecolor{dialinecolor}{rgb}{0.000000, 0.000000, 0.000000}
\pgfsetstrokecolor{dialinecolor}
\node[anchor=west] at (20.100000\du,28.335000\du){$x_1$};
\definecolor{dialinecolor}{rgb}{0.000000, 0.000000, 0.000000}
\pgfsetstrokecolor{dialinecolor}
\node[anchor=west] at (28.950000\du,28.335000\du){$x_2$};
\definecolor{dialinecolor}{rgb}{0.000000, 0.000000, 0.000000}
\pgfsetstrokecolor{dialinecolor}
\node[anchor=west] at (33.256100\du,11.400000\du){$\log\frac{\pi_j(x)}{\pi_{j + 1}(x)}$};
\pgfsetlinewidth{0.050000\du}
\pgfsetdash{{1.000000\du}{1.000000\du}}{0\du}
\pgfsetdash{{0.300000\du}{0.300000\du}}{0\du}
\pgfsetbuttcap
{
\definecolor{dialinecolor}{rgb}{0.000000, 0.000000, 0.000000}
\pgfsetfillcolor{dialinecolor}
\definecolor{dialinecolor}{rgb}{0.000000, 0.000000, 0.000000}
\pgfsetstrokecolor{dialinecolor}
\draw (20.297700\du,18.150000\du)--(29.297700\du,18.150000\du);
}
\pgfsetlinewidth{0.050000\du}
\pgfsetdash{{0.300000\du}{0.300000\du}}{0\du}
\pgfsetdash{{0.300000\du}{0.300000\du}}{0\du}
\pgfsetbuttcap
{
\definecolor{dialinecolor}{rgb}{0.000000, 0.000000, 0.000000}
\pgfsetfillcolor{dialinecolor}
\definecolor{dialinecolor}{rgb}{0.000000, 0.000000, 0.000000}
\pgfsetstrokecolor{dialinecolor}
\draw (29.347700\du,14.100000\du)--(29.297700\du,18.200000\du);
}
\definecolor{dialinecolor}{rgb}{0.000000, 0.000000, 0.000000}
\pgfsetstrokecolor{dialinecolor}
\node[anchor=west] at (29.400000\du,15.700000\du){$\beta (x_2 - x_1)$};
\pgfsetlinewidth{0.050000\du}
\pgfsetdash{}{0pt}
\pgfsetdash{}{0pt}
\pgfsetbuttcap
{
\definecolor{dialinecolor}{rgb}{0.000000, 0.000000, 0.000000}
\pgfsetfillcolor{dialinecolor}
\definecolor{dialinecolor}{rgb}{0.000000, 0.000000, 0.000000}
\pgfsetstrokecolor{dialinecolor}
\draw (19.183100\du,18.688100\du)--(33.600000\du,12.118100\du);
}
\definecolor{dialinecolor}{rgb}{0.000000, 0.000000, 0.000000}
\pgfsetstrokecolor{dialinecolor}
\node[anchor=west] at (33.156200\du,18.30000\du){$\log\frac{\pi_k(x)}{\pi_{k + 1}(x)}$};
\pgfsetlinewidth{0.050000\du}
\pgfsetdash{{1.000000\du}{1.000000\du}}{0\du}
\pgfsetdash{{0.300000\du}{0.300000\du}}{0\du}
\pgfsetbuttcap
{
\definecolor{dialinecolor}{rgb}{0.000000, 0.000000, 0.000000}
\pgfsetfillcolor{dialinecolor}
\definecolor{dialinecolor}{rgb}{0.000000, 0.000000, 0.000000}
\pgfsetstrokecolor{dialinecolor}
\draw (20.297700\du,24.920000\du)--(29.297700\du,24.920000\du);
}
\pgfsetlinewidth{0.050000\du}
\pgfsetdash{{0.300000\du}{0.300000\du}}{0\du}
\pgfsetdash{{0.300000\du}{0.300000\du}}{0\du}
\pgfsetbuttcap
{
\definecolor{dialinecolor}{rgb}{0.000000, 0.000000, 0.000000}
\pgfsetfillcolor{dialinecolor}
\definecolor{dialinecolor}{rgb}{0.000000, 0.000000, 0.000000}
\pgfsetstrokecolor{dialinecolor}
\draw (29.347700\du,20.870000\du)--(29.297700\du,24.970000\du);
}
\definecolor{dialinecolor}{rgb}{0.000000, 0.000000, 0.000000}
\pgfsetstrokecolor{dialinecolor}
\node[anchor=west] at (29.400000\du,22.470000\du){$\beta (x_2 - x_1)$};
\pgfsetlinewidth{0.050000\du}
\pgfsetdash{}{0pt}
\pgfsetdash{}{0pt}
\pgfsetbuttcap
{
\definecolor{dialinecolor}{rgb}{0.000000, 0.000000, 0.000000}
\pgfsetfillcolor{dialinecolor}
\definecolor{dialinecolor}{rgb}{0.000000, 0.000000, 0.000000}
\pgfsetstrokecolor{dialinecolor}
\draw (19.183100\du,25.458100\du)--(33.600100\du,18.888100\du);
}
\end{tikzpicture}